# Anisotropic magnetoresistance and spin polarization of $La_{0.7}Sr_{0.3}MnO_3$ / $SrTiO_3$ superlattices

L. M. Wang

Department of Electrical Engineering, Da-Yeh University, Chang-Hwa 515, Taiwan, R. O. C.

The crystalline structure, anisotropic magnetoresistance (AMR), and magnetization of $La_{0.7}Sr_{0.3}MnO_3$/$SrTiO_3$ (LSMO/STO) superlattices grown by an rf sputtering system are systematically analyzed to study the spin polarization of manganite at interfaces. A perfectly epitaxial growth with sharp interfaces between LSMO and STO layers is confirmed by the transmission electron microscopy (TEM) image and the x-ray diffraction. The presence of positive low-temperature AMR in LSMO/STO superlattices with thinner LSMO layers or thicker STO layers implies that two bands of majority and minority character contribute to the transport properties, leading to a reduced spin polarization. Furthermore, the magnetization of superlattices follows the $T^{3/2}$ law at low temperatures and decays more quickly as the thickness ratio $d_{STO}/d_{LSMO}$ increases, corresponding to a reduced exchange coupling. The results clearly show that the spin polarization is strongly correlated with the influence of interface-induced strain on the structure.

PACS numbers: 75.47.Gk, 68.37.-d, 75.70.Cn

Manganites of the type $R_{1-x}A_xMnO_3$ (R = rare earth, A = Ca, Sr, Ba, and Pb) are considered to be half-metallic and therefore ideal candidates for the use in spin-electronic devices. Recently, a tunneling magnetoresistance (TMR) ratio of more than 1800% at 4 K for $La_{2/3}Sr_{1/3}MnO_3$ /$SrTiO_3$ /$La_{2/3}Sr_{1/3}MnO_3$ trilayer junctions was obtained,[2] leading to an inferred electrode spin polarization of at least 95%. However the reported low-field magnetoresistance for these manganite-based devices decreases rapidly with increased temperature, and even vanishes at temperatures well below the Curie temperature of bulk manganites. It is generally believed that tunneling is a mechanism occurring near the electrode/barrier interface, and the TMR is dominated by charge carriers near the interface boundary.[3, 4] In particular, the rapid decrease of the spin polarization at interfaces with increasing temperature would thus limit the application of the half-metallic materials for spin-electronics devices. That points to an important issue in the physics of their interface properties. Ferromagnetic manganite-insulator superlattices, containing many interfaces of interest, offer the possibility to probe the properties of ultrathin manganite layers and the interface magnetism. Up to now, several groups have reported on the fundamental properties of $La_{1-x}A_xMnO_3/SrTiO_3$ (A= Ca, Sr, and Ba) superlattices.[5-9] For example, Sahana *et al.*[5] and Dörr *et al.*[6] have studied the magnetic and transport properties of La$_{0.7}$Sr$_{0.3}$MnO$_3$/SrTiO$_3$ (LSMO/STO) superlattices. They observed the suppression in both T$_C$ and magnetization accompanying an increase of resistivity as the thickness of LSMO layers decreases. An enhanced high-field magnetoresistance implicated in the magnetically disordered interfaces for La$_{0.7}$Ca$_{0.3}$MnO$_3$/STO superlattices was reported by Jo *et al.*[7] Various magneto-transport properties of manganite-insulator superlattices have been attributed to the strain effect,[8] or interlayer coupling.[9] However, for the ferromagnetic manganite-based superlattices, surprisingly few studies have so far been made on the central problem of the spin polarization in ultrathin manganite layers. This is a key issue for the fabrication of any spin-electronic device composed of manganite/insulator tunneling interfaces.

In this paper we report on the anisotropic magnetoresistance (AMR) and the low-temperature magnetization of LSMO/STO superlattices grown on (001) LaAlO$_3$ substrates. The AMR, being a property determined by the contributions of majority and minority current, can offer information on the spin-dependent band structure.[10] On the other hand, the temperature-dependent magnetization of ferromagnetic superlattices will reflect the temperature dependence of spin polarization near the interfaces. A reduced-spin polarization and a rapidly decreased magnetization with an increase of temperature for superlattices with thinner LSMO layers or thicker STO layers are reported and discussed.

LSMO/STO superlattices grown on (001) LaAlO$_3$ substrate were prepared in an rf magnetron sputtering system as previously described.[11] A buffer STO layer of 60 nm in thickness was deposited prior to the growth of LSMO/STO superlattices to diminish the substrate-induced strain. LSMO and STO layers were alternatively deposited at 700$^o$C in 300-mTorr sputtering gas (Ar and O$_2$, 3:7) until the desired thickness of a superlattice was reached. Sharp interfaces and uniformly continuous layers in superlattices were confirmed by transmission electron microscopy (TEM). The epitaxial growth of superlattices was characterized by an x-ray *θ-2θ* diffractometer using Cu-K$_\alpha$ radiations. For transport measurements, films were photo-lithographically patterned to a 100-μm long by 50-μm wide bridge. The resistivities and magnetizations of superlattices were measured by the standard four-terminal method, and by a superconducting quantum interference device magnetometer, respectively.

Figure 1 is a high-resolution cross-sectional TEM lattice image of a LSMO/STO superlattice denoted by (76/56)$_{12}$ in [010] direction, where the numbers in parentheses correspond respectively to the thicknesses of LSMO and STO layers, and the subscript denotes the total repeated number of bilayers. This figure shows a perfectly epitaxial growth with sharp interfaces between LSMO and STO within less than 1 nm. Furthermore, the TEM analysis shows that the superlattice maintains the in-plane crystal coherency at the interfaces with an in-plane lattice constant of 3.849 Å, which ranges just between the lattice constant of 3.79 Å for the LaAlO$_3$ and the pseudocubic lattice constant of 3.876 Å for bulk LSMO.[12] This indicates that the LSMO and STO (*a* axis = 3.905 Å) layers are in an in-plane compressive stress state due to a smaller lattice constant of substrate.[8] Normally along the film there can be seen a slightly incoherent growth with out-of-plane lattice constants of 3.939 and 3.896 Å in LSMO and STO layers, respectively. This corresponds to an elongation of the *c* axis by 0.063 Å for LSMO and a negligible *c*-axis strain in STO layers. The presented TEM image shows a more clearly coherent hetrostructure with sharp interfaces than that reported on LSMO/STO superlattices grown by pulsed laser deposition.[6] It also demonstrates that high-quality perovskite superlattices can be achieved by a lower-cost sputtering technique. Figure 2(a) shows the x-ray θ-2θ diffraction spectra in the region near the (002) peak for a series of LSMO/STO superlattices. Clearly, the 2θ position of the (002) peak, which corresponds to the out-of-plane lattice constant, is dependent on the layer thickness in LSMO/STO superlattices, indicating a variation of out-of-plane strain among these superlattices. The observed average *c*-axis lattice constant of 3.903 Å for the (76/56)$_{12}$ superlattice is close to that seen in the TEM image. For the (142/82)$_6$ and (76/147)$_{12}$ superlattices, the determined average *c*-axis lattice constants are 3.879 Å and 3.911 Å respectively.

To put it briefly, for a superlattice with thicker STO layers or thinner LSMO layers, an enhanced *c*-axis lattice was observed. This feature is similar to that observed in LSMO/STO superlattices grown on STO substrates.[5] Figure 2(b) shows the x-ray scan for sample $(76/147)_{12}$, in which only the (001) and (002) diffraction peaks of film were observed, also indicating that the film has growth that is oriented along the *c*-axis. Furthermore, the in-plane orientation of the superlattices has been studied by the x-ray $\Phi$-scanning on the orthorhombic (332) LSMO diffraction peak. The inset of Fig. 2(b) gives a typical pattern for $(76/56)_{12}$ superlattice. As can be seen, the fourfold symmetry of this pattern clearly indicates the in-plane epitaxial structure. In Fig. 2(a), the presence of the satellite peaks on both sides of the main peak (002) confirms that a periodic structure in the superlattices has been achieved. The modulation wavelength, $\Lambda = d_{LSMO} + d_{STO}$, where $d_{LSMO}$ and $d_{STO}$ are the thickness of the LSMO and STO layers respectively, can be calculated from the separation of two successive peaks (i and i +1) using the equation: $\Lambda = (\lambda/2)[1/(\sin\theta_i - \sin\theta_{i+1})]$, where $\lambda$ is the x-ray wavelength ($\lambda$ = 1.5406 Å). The modulation wavelength $\Lambda$ = 128 Å obtained from the x-ray data for $(76/56)_{12}$ superlattice is in close agreement with that of 132 Å observed in the TEM image.

Figure 3 shows the zero-field resistivities $\rho$ as a function of temperature for a series of LSMO/STO superlattices and the LSMO film. The inset of Fig. 3 illustrates the temperature dependence of the magnetoresistance ratio, defined as MR(7 T) = $[\rho(H = 7\ T) – \rho(H = 0)]/\rho(H = 7\ T)$, for the corresponding samples. It can be seen that with decreasing $d_{LSMO}$ or increasing $d_{STO}$, the value of $\rho$ increases, and $\rho$ reveals a metallic state at low temperatures for all superlattices. Also visible in the inset of Fig. 3 is that a maximum MR(7 T) value occurs near the Curie temperature for all the superlattices and the 800-Å LSMO film. The maximum MR values for LSMO/STO superlattices occurring at temperatures 345 -350 K are in the range of 37 – 38%, which is slightly smaller than that of 41% at 352 K for the LSMO film. Figure 4 shows the temperature dependence of AMR for a series of LSMO/STO superlattices. The AMR is defined as AMR(7 T) = MR(7 T)$_{//}$ - MR(7 T)$_{\perp}$, where MR(7 T)$_{//}$ and MR(7 T)$_{\perp}$ denote the longitudinal (H // the electric current) and transverse (H $\perp$ the electric current) magnetoresistance ratio, respectively. Here the data were obtained with currents along crystal [100] direction, and magnetic fields applied in the film plane to eliminate the demagnetization effect. It can be observed that the AMR behavior is markedly different among the superlattices with different thicknesses of LSMO or STO layers.  For example, the AMR of (76/56)$_{12}$ superlattice shows a negative value of -2% at low temperatures that gradually diminishes at higher temperatures; whereas the AMR for (76/147)$_{12}$ superlattice, a sample with thicker STO layers, shows a positive value of 3.4% at low temperatures. It is found that the low-temperature AMR value increases gradually and changes to a positive value for superlattices with thinner LSMO layers or thicker STO layers. The top inset of Fig. 4 shows the AMR behavior of the 800-Å LSMO film for comparison. One can find that the AMR of LSMO film at low temperatures is very low, around -0.06%, then gradually drops to the lowest point of -8.9% at the highest temperature measured. This presents a very similar feature compared to that of $La_{0.7}Ca_{0.3}MnO_3$ films observed by Ziese,[10] but a quit different behavior from that of the superlattices studied. The excess negative AMR near $T_C$ for high-quality manganite films has been suggested to be related to the inhomogeneous magnetic state.[13] Moreover, contrary to the high-quality epitaxial films, the AMR, being highest at low temperatures with no clear peak near $T_C$, was also observed in a polycrystalline $La_{0.7}Ca_{0.3}MnO_3$ film.[10] Since the grain-boundary or interface-scattering resistance is isotropic and will be canceled in the calculation of the AMR, the anisotropic transport properties have been suggested to be related to the local lattice distortions of the Mn-O bonds.[13] Thus, the AMR reflects an intrinsic transport property and is related to the crystalline structure. This gives a good account for the variegated AMR behaviors observed in the LSMO/STO superlattices. It must be recalled here that a variation of out-of-plane strain can be deduced from the x-ray results, as previously mentioned. Turning now to concentrate on the low-temperature AMR features of LSMO/STO superlattices, Ziese has recently analyzed the AMR behaviors of $La_{0.7}Ca_{0.3}MnO_3$ and $Fe_3O_4$ films within the two-current model and an atomic $d$-state calculation.[10] For the $La_{0.7}Ca_{0.3}MnO_3$ film, in which a negative AMR was observed, he pointed out a weak influence of a minority spin band. Additionally, based on the observation of a positive AMR in the $Fe_3O_4$ film, it was suggested that at least two bands of majority and minority character contribute to the transport properties of $Fe_3O_4$, leading to its lower spin polarization. The same is true of the presented case for LSMO/STO superlattices. The presence of positive AMR in LSMO/STO superlattices with thinner LSMO layers or thicker STO layers implies a reduced spin polarization occurring in them. The lower inset of Fig. 4 shows the hysteretic AMR curve associated with the normalized magnetization $M/M(H = 5 kG)$ of the $(76/82)_{12}$ superlattices performed at 5 K. It can be seen that the AMR remains nearly constant as the magnetization saturates, while an anomalous AMR behavior appears at low fields. Considering that the magnetization is not fully aligned in the low-field region, the low-field AMR is supposed to contain an extrinsic contribution originating from the grain-boundary or interface magnetization anisotropy.[14] The nearly field-independent AMR observed in the high-field region also indicates that the Lorentz force effect is suppressed in the LSMO/STO superlattices with increasing the spacer layer thickness. It is known the Lorentz force bends the trajectory of the carriers and then contributes a positive MR at high fields for the transverse configuration. If the Lorentz mechanism dominates the AMR at high fields, it may be expected that the AMR value will be negative and decrease with increasing fields according to the definition of AMR = $MR_{//}$ - $MR_{\perp}$. In fact, a marked Lorentz-force-like contribution is observed only on the high-field AMR of the LSMO film (not shown). The results give evidence that the AMR behavior is controllable by the artificial superlattices.

To explore the temperature dependence of spin polarization P(T), a convenient method is to study the temperature dependence of magnetization M(T).[15, 16] The measurement of M(T) on LSMO/STO superlattices is supposed to reflect the P(T) over the ultrathin LSMO layers. The M(T), being proportional to P(T), should follow the $T^{3/2}$ law according to M(T)/M(0) = 1- $CT^{3/2}$ = 1- $kC_{bulk}T^{3/2}$, where $C_{bulk}$ is the constant describing the decrease of the bulk magnetization due to thermal excitation of the spin wave, and k = 2 for the ideal surface case.[17] Figures 5(a) and 5(b) show the normalized magnetization M/M(5 K) measured with H = 500 G as a function of $T^{3/2}$ for two series of superlattices with varied thicknesses of STO layers and LSMO layers respectively. It can be seen that the M(T) follows the $T^{3/2}$ law at temperatures below 140 K for all the superlattices and the LSMO film. It can also be seen that the M(T) of superlattices decays more quickly with increasing temperature, corresponding to a fast decay of P(T), as the STO layers become thicker or the LSMO layers become thinner.

It is found that the value of constant C increases with an increase of $d_{STO}/d_{LSMO}$ ratio, as shown in the inset of Fig. 5(a). Taking k = 1 for the LSMO film, the parameter k of superlattices also shows a monotonous increase with an increase of $d_{STO}/d_{LSMO}$ ratio, as seen in the inset of Fig. 5(b). The presented values of k, in the range of 1.5 – 3.0, are comparable to those in the range of 1.1 – 4.2 derived from the P(T) of $La_{2/3}Sr_{1/3}MnO_3$ TMR devices.[16] It is noteworthy that the parameter k can be a measure of the exchange coupling $J_\perp$ on a path perpendicular to the interfaces.[15] According to Mathon,[18] the value of k = 3.0 for the $(76/147)_{12}$ superlattice corresponds to $J_\perp/J \approx 0.3$, where J is the exchange interaction in the bulk. It is known that a reduced exchange coupling should not completely separate the majority carrier conduction band from the minority band, leading to an incomplete polarization of the carriers.[19] This result is consistent with the inference from the AMR properties previously discussed.

In summary, high-quality LSMO/STO superlattices have been grown in an rf sputtering system, and characterized by the TEM image, x-ray diffraction, and resistive measurement. These superlattices offer a good opportunity to probe the properties of ultrathin manganite layers and the interface magnetism. According to Ziese,[10] the presence of positive low-temperature AMR in LSMO/STO superlattices with thinner LSMO layers or thicker STO layers implies that at least two bands, of majority and minority character, contribute to the transport properties, leading to a reduced spin polarization. Furthermore, the M(T) of superlattices follows the $T^{3/2}$ law at low temperatures and decays more quickly as the $d_{STO}/d_{LSMO}$ ratio increases. The results clearly show that the spin polarization is strongly correlated with the influence of interface-induced strain on the structure. The interface-induced strain must be taken into account in the fabrications of TMR devices based on doped manganites.

The author thanks the National Science Council of the Republic of China for financial support under Grant Nos. NSC 93-2112-M-212-001 and NSC 94-2623-7-212-003-AT. This work was also partially supported by Da-Yeh University under Grant No. ORD-9303.

# References

1. E. D. Wollan and W. C. Köhler, Phys. Rev. **100**, 5643 (1955).

2. M. Bowen, M. Bibes, A. Barthélémy, J.-P. Contour, A. Anane, Y. Lemaître, and A. Fert, Appl. Phys. Lett. **82**, 233(2003).

3. J. –H. Park, E. Vescovo, H. –J. Kim, C. Kwon, R. Ramesh, and T. Venkatesan, Phys. Rev. Lett. **81**, 1953(1998).

4. Jagadeesh S. Moodera, Janusz Nowak, and Rene J. M. van de Veerdonk, Phys. Rev. Lett. **80**, 2941(1998).

5. M. Sahana, T. Walter, K. Dörr, K.-H. Müller, D. Eckert, and K. Brand, J. Appl. Phys. **89**, 6834 (2001).

6. K. Dörr, T. Walter, M. Sahana, K.-H. Müller, K. Nenkov, K. Brand, and L. Schultz, J. Appl. Phys. **89**, 6973 (2001).

7. Moon-Ho Jo, Neil D. Mathur, Jan E. Evetts, Mark G. Blamire Manuel Bibes, and Josep Fontcuberta, Appl. Phys. Lett. **75**, 3689 (1999).

8. Yafeng Lu, J. Klein, C. Höfener, B. Wiedenhorst, J. B. Philipp, F. Herbstritt, A. Marx, L. Alff, and R. Gross, Phys. Rev. B **62**, 15806 (2000).

9. M. Izumi, Y. Ogimoto, Y. Okimoto, T. Manako, P. Ahmet, K. Nakajima, T. Chikyow, M. Kawasaki, and Y. Tokura, Phys. Rev. B **64**, 64429 (2001).

10. M. Ziese, Phys. Rev. B **62**, 1044 (2000); M. Ziese and S. P. Sena, J. Phys.: Condens. Matter **10**, 2727 (1998).

11. L. M. Wang, H. H. Sung, B. T. Su, H. C. Yang, and H. E. Horng, J. Appl. Phys. **88**, 4236 (2000).

12. Michael C. Martin, G. Shirane, Y. Endoh, K. Hirota, Y. Moritomo, and Y. Tokura, Phys. Rev. B **53**, 14285 (1996).

13. V. S. Amaral, A. A. C. S. Lourenco, J. P. Araújo, P. B. Tavares, E. Alves, J. B. Sousa, J. M. Vieira, M. F. da Silva, and J. C. Soares, J. Magn. Magn. Mater. **211**, 1 (2000).

14. R. Mathieu, P. Svedlindh, R. A. Chakalov, and Z. G. Ivanov, Phys. Rev. B **62**, 3333 (2000).

15. D. Mauri, D. Scholl, H. C. Siegmann, and E. Kay, Phys. Rev. Lett. **61**, 758 (1988).

16. V. Garcia, M. Bibes, A. Barthélémy, M. Bowen, E. Jacquet, J.-P. Contour, and A. Fert, Rev. B **69**, 052403 (2004).

17. D. Mills and A. Maradudin, J. Phys. Chem. Solids **28**, 1855 (1967).

18. J. Mathon, Physica B **149**, 31 (1988).

19. H. Y. Hwang, S-W. Cheong, N. P. Ong, and B. Batlogg, Phys. Rev. Lett. **77**, 2041 (1996).

**Figure captions**

Figure 1. High-resolution cross-sectional TEM lattice image of a LSMO/STO superlattice.

Figure 2. (a) X-ray $\theta$-$2\theta$ diffraction spectra in the region near the (002) peak for a series of LSMO/STO superlattices. The satellite peaks are indicated by star symbols. (b) Wide-range x-ray scan for the $(76/147)_{12}$ superlattice. The inset shows its x-ray $\phi$-scanning on the orthorhombic (332) LSMO diffraction peak.

Figure 3. Zero-field resistivities $\rho$ as a function of temperature for a series of LSMO/STO superlattices and the LSMO film. The inset illustrates the temperature dependence of magnetoresistance ratio for corresponding samples.

Figure 4. Temperature dependence of AMR for a series of LSMO/STO superlattices. Top inset: the AMR behavior of the 800-Å LSMO film for comparison. Lower inset: the hysteretic AMR curve associated with the normalized magnetization M/M(5 kG) of the $(76/82)_{12}$ superlattices performed at 5 K.

Figure 5. Normalized magnetization M/M(5 K) measured with H = 500 G as a function of $T^{3/2}$ for (a) a series of superlattices with varied thicknesses of STO layers and (b) a series of superlattices with varied thicknesses of LSMO layers. Inset of (a): the value of constant C as a function of $d_{STO}/d_{LSMO}$ ratio. Inset of (b): the parameter k as a function of $d_{STO}/d_{LSMO}$ ratio. The dashed lines are for the purpose of guiding the eye.

Fig. 1

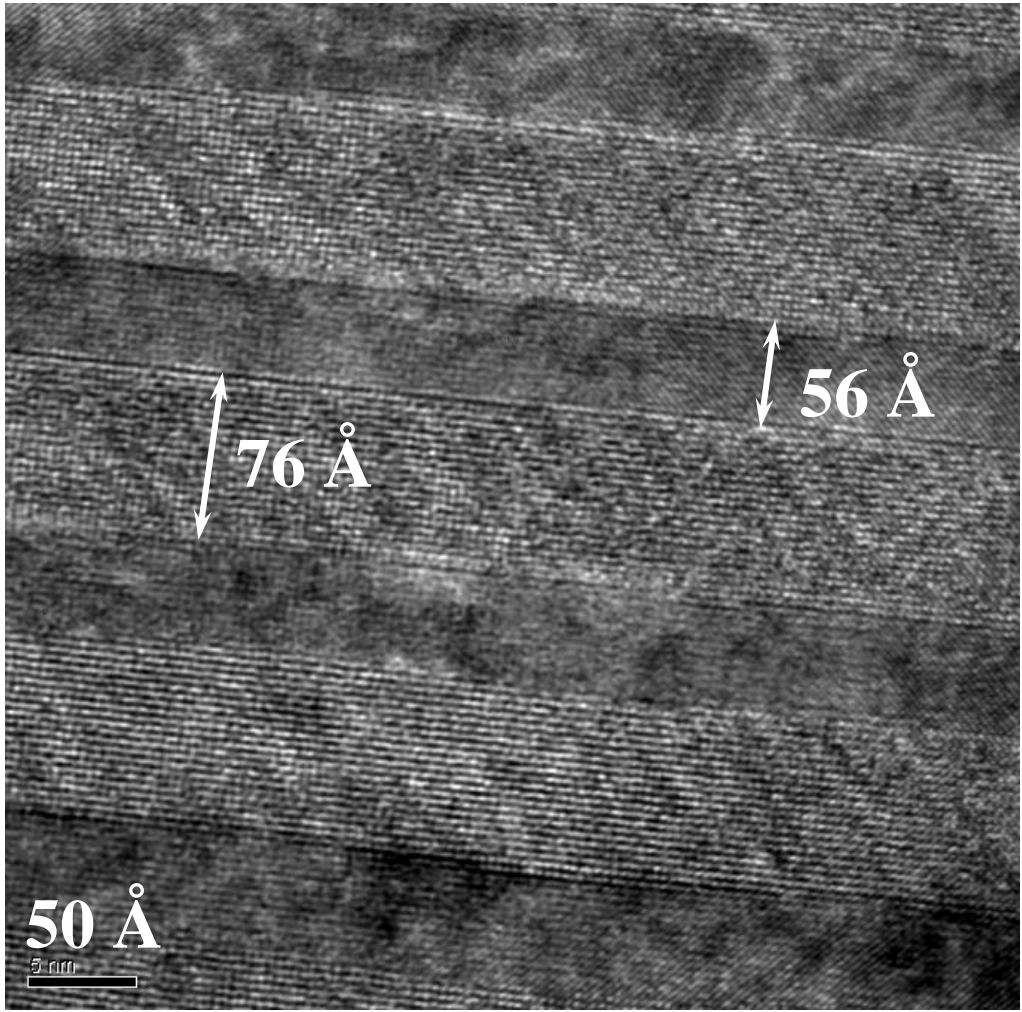

Fig. 2

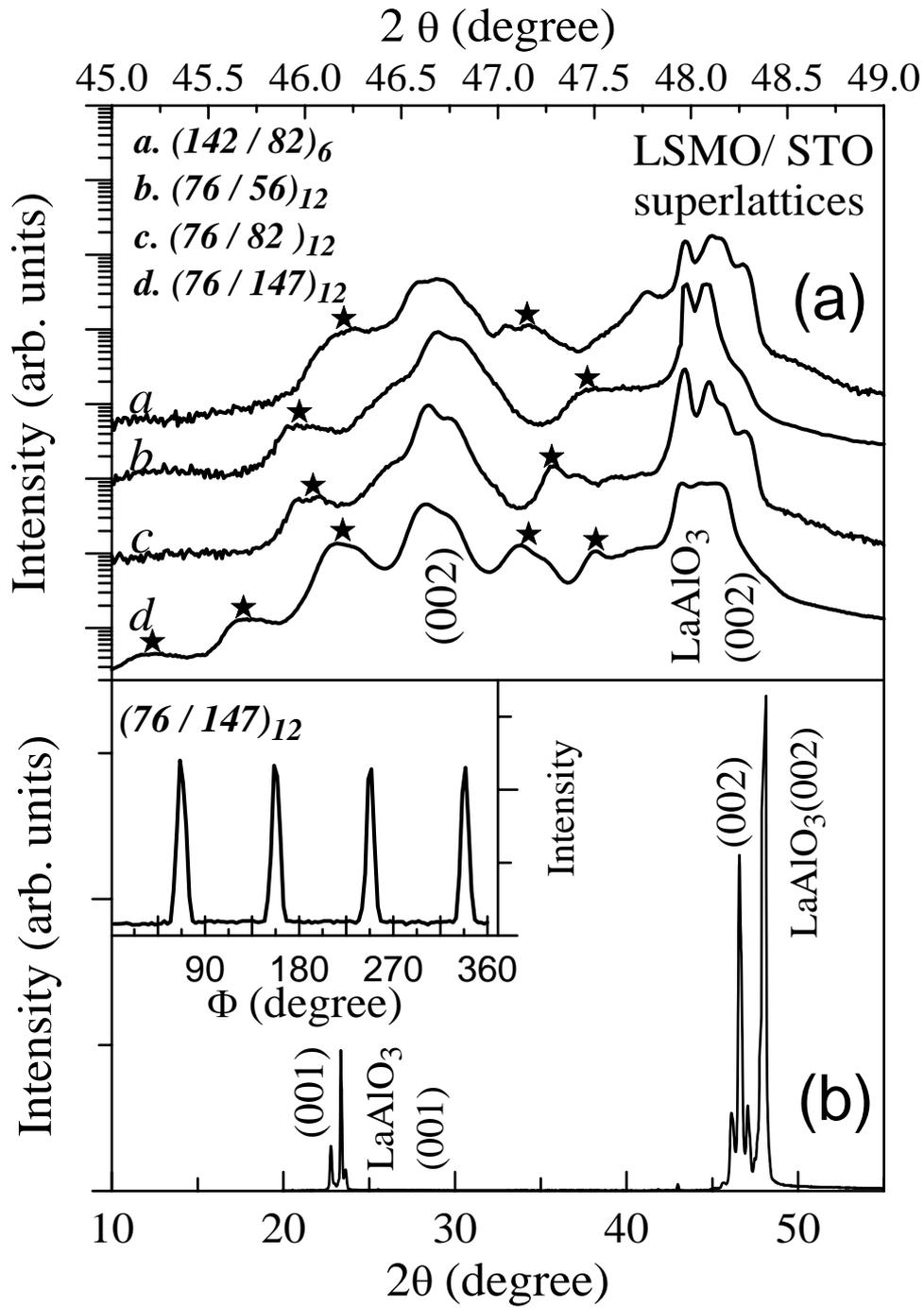

Fig. 3

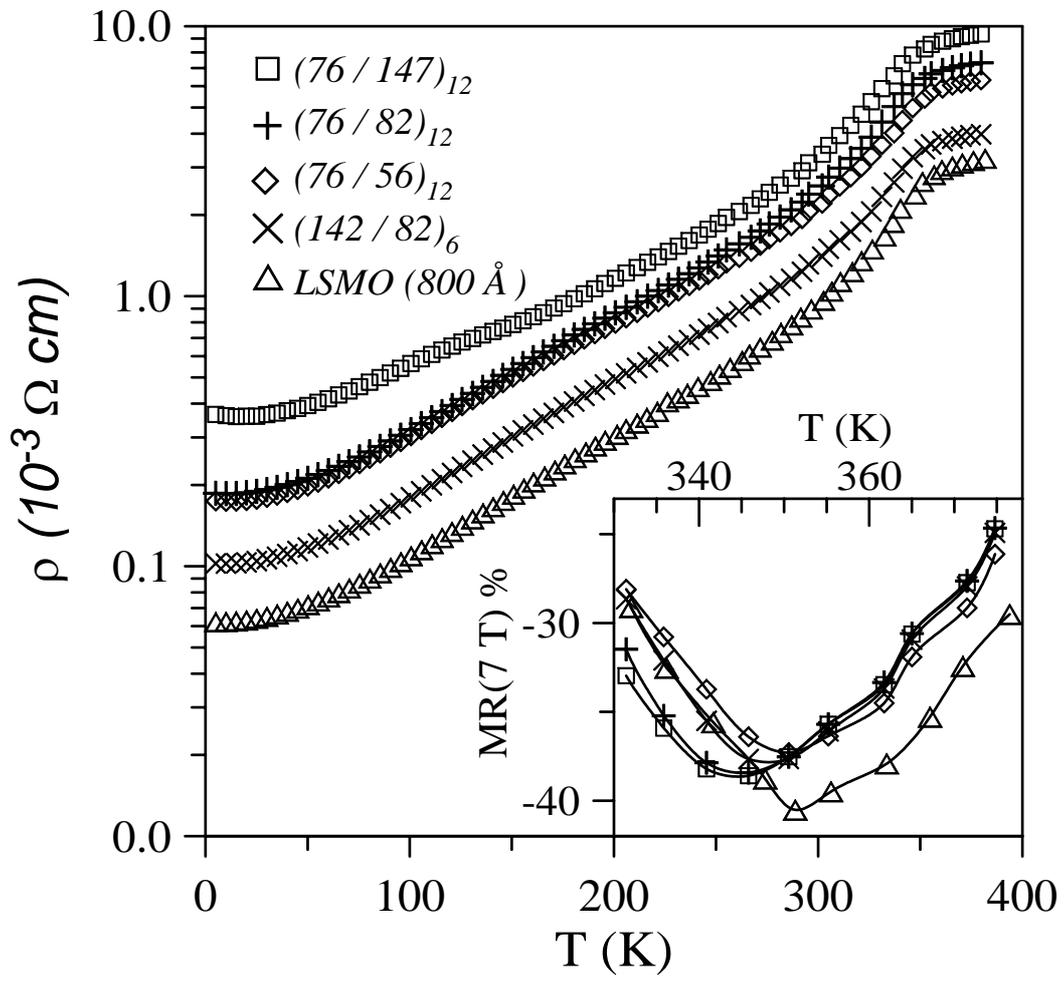

Fig. 4

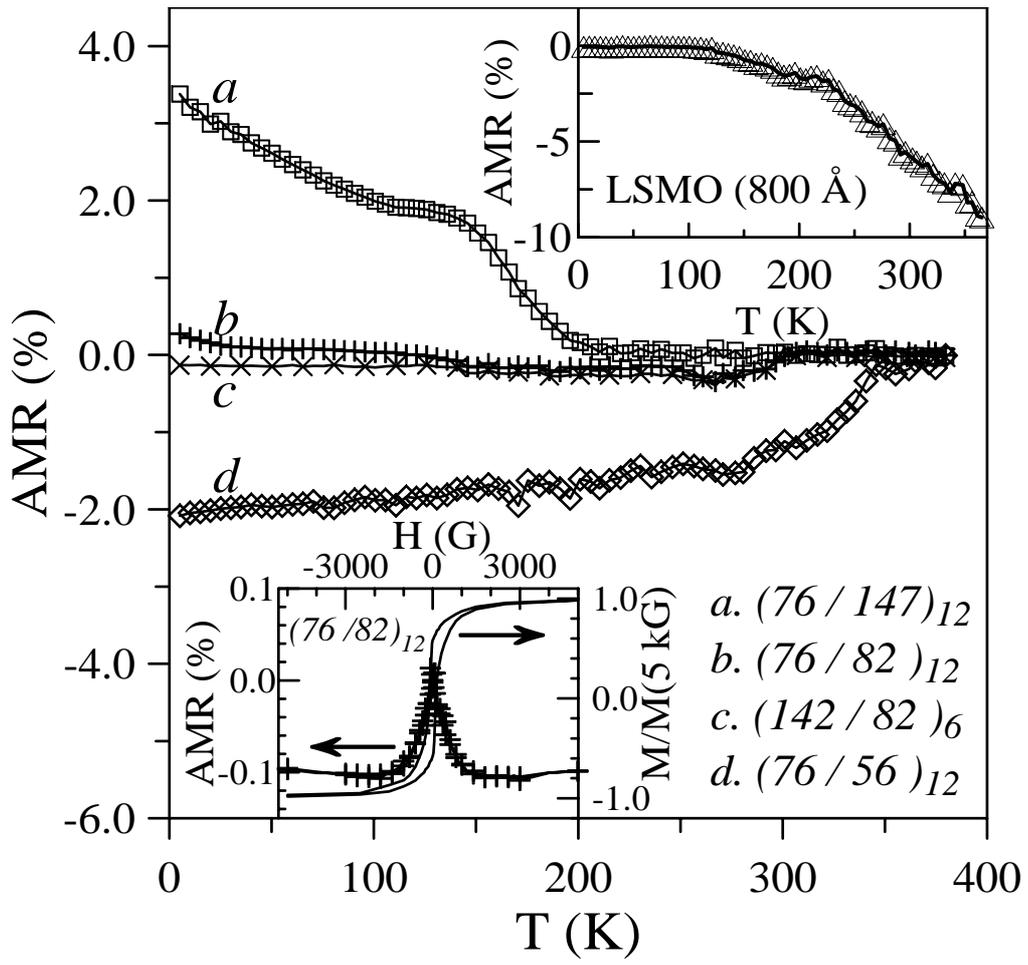

Fig. 5

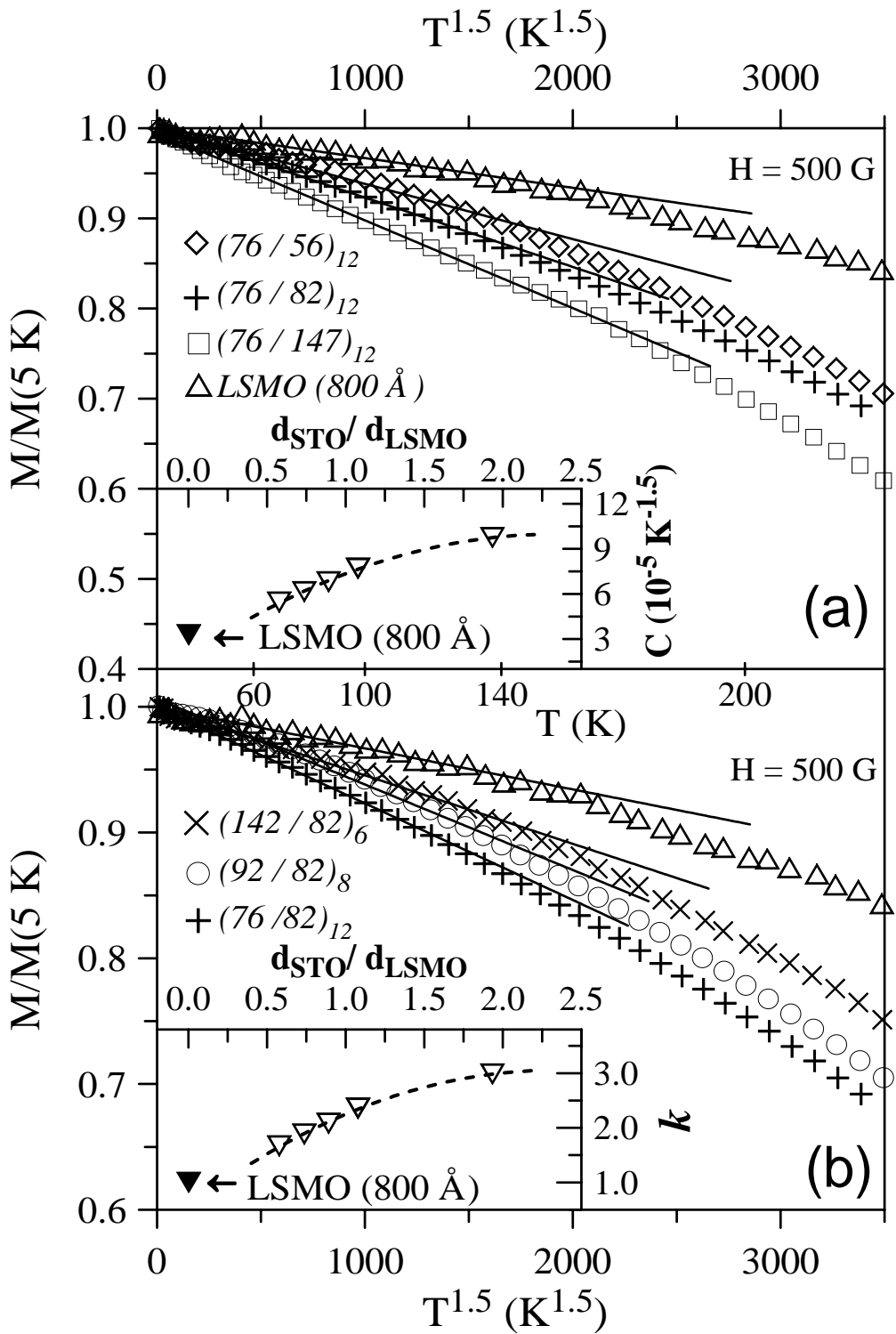